# Atomically imprinted graphene plasmonic cavities

Brian S. Y. Kim[1,2,†,*], Aaron J. Sternbach[1,†], Min Sup Choi[2,†], Zhiyuan Sun[1], Francesco L. Ruta[1,3], Yinming Shao[1], Alexander S. McLeod[1], Lin Xiong[1], Yinan Dong[1], Anjaly Rajendran[2,4], Song Liu[2], Ankur Nipane[4], Sang Hoon Chae[2], Amirali Zangiabadi[3], Xiaodong Xu[5], Andrew J. Millis[1], P. James Schuck[2], Cory. R. Dean[1], James C. Hone[2], D. N. Basov[1,*]

[1]*Department of Physics, Columbia University, New York, NY, USA*
[2]*Department of Mechanical Engineering, Columbia University, New York, NY, USA*
[3]*Department of Applied Physics and Applied Mathematics, Columbia University, New York, NY, USA*
[4]*Department of Electrical Engineering, Columbia University, New York, NY, USA*
[5]*Department of Physics, University of Washington, Seattle, WA, USA*
[†]*These authors contributed equally to this work*

**Plasmon polaritons in van der Waals (vdW) materials hold promise for next-generation photonics[1-4]. The ability to deterministically imprint spatial patterns of high carrier density in cavities and circuitry with nanoscale features underlies future progress in nonlinear nanophotonics[5] and strong light-matter interactions[6]. Here, we demonstrate a general strategy to atomically imprint low-loss graphene plasmonic structures using oxidation-activated charge transfer (OCT). We cover graphene with a monolayer of $WSe_2$, which is subsequently oxidized into high work-function $WO_x$ to activate charge transfer. Nano-infrared imaging reveals low-loss plasmon polaritons at the $WO_x$/graphene interface. We insert $WSe_2$ spacers to precisely control the OCT-induced carrier density and achieve a near-intrinsic quality factor of plasmons. Finally, we imprint canonical plasmonic cavities exhibiting laterally abrupt doping profiles with single-digit nanoscale precision via programmable OCT. Specifically, we demonstrate technologically appealing but elusive plasmonic whispering-gallery resonators based on free-standing graphene encapsulated in $WO_x$. Our results open avenues for novel quantum photonic architectures incorporating two-dimensional materials.**

Graphene-based plasmonic nanostructures offer exciting opportunities for engineered light-matter interaction at the nanoscale[7,8]. Thus far, strategies for realizing plasmonic nanostructures in graphene have largely focused on either physically etching graphene using nanolithography[9-11] or modulating the density profile via meta-structured gate dielectrics[12]. Whereas the first approach yields fabrication-related nanoscale disorder[10], the second pathway creates blurred doping profiles due to electric field fringing[13]. These two effects are particularly detrimental for structures with nanoscale elements, ultimately limiting the device performance[10,11,14].

In this work, we show that charge-transfer doping via a nearby high work-function layer provides a powerful complementary method for engineering plasmonic nanostructures. The band alignment within these tailored heterostructures naturally creates built-in potential across the constituent layers and induces charge transfer[15]. These plasmonic devices operate without the need for gating, allowing for the generation of high carrier density without the risk of electrical breakdown. Furthermore, the charge-transfer process produces laterally abrupt boundaries between doped and charge-neutral regions. This latter attribute allows us to implement plasmonic structures with sharp features and designer doping profiles.

Here, we demonstrate the creation of low-loss graphene plasmonic cavities using OCT. In our approach, charge-neutral graphene is first covered with monolayer $WSe_2$. Subsequent oxidization, using UV-ozone treatment, transforms monolayer $WSe_2$ into monolayer $WO_x$ – a high work-function transition-metal oxide (TMO) (**Fig. 1**a; **Methods**). Large work-function mismatch is known to cause electrons to migrate from graphene to $WO_x$, leaving the graphene strongly hole-doped[16]. This oxidation process is self-limiting. For example, if multilayer $WSe_2$ is exposed to UV-ozone treatment, only the top-most atomic monolayer of $WSe_2$ is oxidized and transformed into $WO_x$, as evidenced by cross-sectional transmission electron microscopy (TEM) images and Raman spectroscopy (**Figs. 1**b–d). Graphene remains intact in our $WO_x$/graphene structures as oxidized $WSe_2$ serves as a high interlayer ozone diffusion barrier[17]. The net result is an atomically engineered pristine $WO_x$/graphene interface devoid of mid-gap interface states or interface dipoles[18], which would otherwise hinder the charge transfer. The resulting high density of Dirac electrons hybridizes with light, producing propagating mid-infrared graphene plasmon polaritons on demand.

We begin with a survey of plasmon polaritons at the $WO_x$/graphene ($WO_x$/G) interface. **Fig. 1**e,f shows data collected at ambient conditions using scattering-type scanning near-field optical microscopy (**Methods**)[19]. Within the $WO_x$/G regions, oscillations of the near-field amplitude *S* are observed. We attribute these characteristic oscillations to plasmonic standing waves[19], confirming the presence of free carriers in $WO_x$/G prompted by OCT. Propagating surface plasmon polaritons are observed in the vicinity of (1) edges and (2) topographic defects, and (3) along one-dimensional channels confined within $WO_x$/G boundaries (**Fig. 1**e,f).

The degree of charge transfer can be precisely controlled by taking advantage of self-limiting oxidation. We fabricated structures with 1-3 layers of pristine $WSe_2$ separating graphene and $WO_x$. The series of images in **Fig. 2**a reveals a systematic decrease in the spatial period of the plasmonic fringes with increased WSe$_2$ spacer thickness. These fringes are a product of plasmons launched by the tip that travel to the sample edge before returning to the tip. Round-trip plasmons display $\lambda_p/2$ periodicity, where $\lambda_p$ is the plasmon wavelength[19]. We also observe plasmons emanating from the sample edge having $\lambda_p$ periodicity. In graphene, the magnitude of $\lambda_p$ allows one to read the carrier density $n$ directly off plasmonic images since $\lambda_p \sim \sqrt{n}$[19]. The analysis of data in **Fig. 2** reveals that the strength of an OCT process is systematically varied by adjusting the number of intact WSe$_2$ layers separating graphene and WO$_x$. Note that back-gating can also independently modulate the plasmonic response of our devices, as indicated by the systematic tuning of $\lambda_p$ using the SiO$_2$/Si back-gate voltage (**Fig. S**1).

In **Fig. 2**b, we plot graphene Fermi energy $E_F$ vs WSe$_2$ spacer thickness $t$, where $E_F$ is obtained from the relation $E_F = \hbar v_F \sqrt{\pi n}$[20], where $\hbar$ is the reduced Planck's constant and $v_F$ is the Fermi velocity (see **Supplementary Discussion** 1 and **Fig. S**2 for quantitative extraction of $n$ and $E_F$ from the plasmon fringes). The attained $E_F$ approaches ~670 meV in WO$_x$/G and can be adjusted to ~400 meV with a 3-layer WSe$_2$ spacer. These values agree well with $E_F$ extracted using Kelvin probe force microscopy (KFPM), corroborating the control of $E_F$ by using WSe$_2$ spacers (**Figs. 2**b,**S**3,**S**4). This control can be understood in terms of the electrostatic band alignment of WO$_x$/G expressed using the work-function mismatch between graphene and WO$_x$ as:

$$\Phi_{WOx} = \Phi_G + nt/\varepsilon, \qquad (1)$$

where $\Phi_{WOx}$ ($\Phi_G$) is the work-function of WO$_x$ (graphene) and $\varepsilon$ is the dielectric constant of WSe$_2$ (**Fig. 2**c,**S**5; **Supplementary Discussion** 2)[16]. Here, we use $\varepsilon = 7.2$ for WSe$_2$[20], and $\Phi_G$ = 4.6 eV + $E_F$. The second term on the right-hand side of equation (1) is simply the potential drop developed across insulating WSe$_2$ due to charge transfer, derived from Poisson's equation.

The dependence of $E_F$ on $t$ is captured by equation (1) with $\Phi_{WOx}$ as a fitting parameter (**Fig. 2**b). The extracted value for $\Phi_{WOx}$ of ~5.6 eV also agrees with the accepted value for non-stoichiometric $WO_x$[21]. Notably, graphene becomes doped even when the $WSe_2$ spacer is replaced with hBN. This latter finding indicates that OCT operates for various insulating spacers as long as the graphene Dirac point lies deep within the spacer band gap (**Figs. 2**b,**S**6; **Supplementary Discussion** 2). Thus, our nano-imaging data verify that the observed doping originates from work-function mismatch.

We next show that the OCT preserves the high quality factor $Q$ of plasmon polaritons in $WO_x/WSe_2/G$ structures. It is customary to define $Q = q_P'/q_P''$, where $q_P'$ and $q_P''$ are the real and imaginary components of the complex plasmonic momentum $q_P = q_P' + iq_P''$. We obtain $q_P'$ and $q_P''$ from the observed $\lambda_P$ and plasmon propagation length $L$, using the standard relations $q_P' = 2\pi/\lambda_P$ and $L = 1/2q_P''$[19,22]. $Q$ reaches ~16 on $WO_x/G$ (**Fig. 2**d), thus surpassing $Q$ of ~5 in bare graphene microcrystals on $SiO_2$[23]. With the addition of $WSe_2$ spacers, $Q$ is enhanced beyond 20, which is on par with that of low-loss plasmons observed in hBN-encapsulated high-mobility graphene at room temperature[19,22]. We quantitatively analyzed the origin of the exceptional $Q$ by considering the three fundamental loss channels: dielectric losses from hBN, intrinsic thermal phonons in graphene, and charged impurities in $WO_x$ (**Supplementary Discussion** 3). Our analysis shows that these three loss channels largely account for the observed $Q$ (**Fig. 2**e). In particular, we extract a remarkably low charged impurity density of ~$2.3\times10^{11}$ cm$^{-2}$ for $WO_x$. The associated charged-impurity scattering rate $\gamma_{imp}$ therefore falls below that for intrinsic phonons in graphene[19,22] even in the absence of $WSe_2$ spacers. The totality of our observations indicates that our OCT method produces a low-loss interface suitable for supporting plasmon polaritons with long-range propagation.

The rapid enhancement of $Q$ with $WSe_2$ spacers can be attributed to the characteristic exponential dependence of $\gamma_{imp}$ on the distance $l$ between charged impurities in $WO_x$ and graphene[24]. $\gamma_{imp}$ can be parametrized by the decay length $l_{imp}$ as $\gamma_{imp} = \gamma_{imp,0}\times\exp(-l/l_{imp})$, where $\gamma_{imp,0}$ is $\gamma_{imp}$ in the limit where charged impurities reside within the graphene lattice sites (**Fig. S7**). The extracted $l_{imp}$ is ~5 Å, which indicates that even monolayer $WSe_2$ (~6 Å) can effectively shield losses from charged impurities. Plasmon polaritons residing in WO$_x$/1L-WSe$_2$/G structures, therefore, perform at a near-intrinsic level at room temperature, essentially being limited by thermal phonons in graphene.

We now turn to plasmonic cavities produced by programmable OCT. To achieve doping in spatially precise patterns, we transferred a pre-etched hBN mask onto WSe$_2$/graphene/hBN heterostructures (**Fig. 3**a,b). The OCT approach then projects laterally abrupt carrier density profiles following the patterns defined in the hBN mask. We stress that the doped and charge-neutral graphene monolayer regions remain fully protected by a combination of hBN and WO$_x$ layers during the imprinting process. As a result, the observed $E_F$ and $Q$ are identical to those obtained in unpatterned structures (**Fig. S**8), indicating the pristine condition of the active plasmonic layers and engineered electrostatic edges in our imprinted structures.

Notably, cavities in **Fig. 3**c show clear plasmonic fringes arising from the reflection of plasmons at the engineered electrostatic edges within the continuous graphene layer. Our simulations show that plasmon reflection at electrostatic edges is rapidly suppressed by nanometer-scale blurring of the density profile and vanishes with a blurring of only ~6 nm (**Fig. S**9). We therefore deduce that OCT produces laterally abrupt doping profiles with previously unattainable single-digit nanoscale precision. In comparison, engineering such sharp electrostatic edges is challenging in gated structures due to electric field fringing in the gate dielectric[13]. We remark that reflections of plasmons at the edges of gated regions in graphene have not been previously observed.

Next, we image emergent plasmonic effects in free-standing graphene cavities encapsulated in WO$_x$. A characteristic feature of these cavities is the enhanced density as charge transfer is simultaneously prompted by the two proximal WO$_x$ surfaces (**Fig. 4**a,b). In these latter structures, the top WO$_x$ covers the entire free-standing graphene, whereas the bottom WO$_x$ covers only the targeted regions on the same graphene defined by the bottom hBN mask. The net result is highly doped freestanding WO$_x$/G/WO$_x$ plasmonic cavities surrounded by moderately doped WO$_x$/G structures. We find that $E_F$ in the WO$_x$-encapsulated regions is ~750 meV, corresponding to a roughly 30% enhancement in hole density. This finding is in good correspondence with the electrostatic model discussed above (**Supplementary Discussion** 2; **Fig. S**10). The combination of higher density and reduced dielectric screening in free-standing structures results in a $\lambda_p$ as long as ~1 μm[25].

When cavity dimensions are reduced down to ~$\lambda_p$, the real-space plasmonic patterns become governed by radially confined standing waves along the circumference of the cavity (**Fig. 4**c). These salient features are plasmonic versions of whispering-gallery modes (WGM)[5] – edge plasmons looping around the cavity due to continuous total internal reflection. **Figure 4**d shows the simulated $z$-component of the electric field $E_z$ solved for eigenmodes bound within the cavity (**Methods**). The mode character is governed by the excitation laser frequency $\omega$, switching from quadrupolar at 940 cm$^{-1}$ to hexapolar at 1080 cm$^{-1}$. The general agreement of our observation with these simulations corroborates the presence of plasmonic WGMs in our cavities. Note that the standing-wave modes with cylindrical symmetry are also present in our cavities in **Fig. 4**c. But these latter cylindrical modes give rise to a single peak near the cavity center rather than multiple oscillatory fringes since $\lambda_p$ is comparable to the cavity dimensions (**Figs. S**11,**S**12).

We stress that our novel low-loss free-standing photonic platform based on OCT enables the implementation of technologically significant yet elusive plasmonic whispering-gallery cavities in graphene. The appeal of plasmonic WGMs in graphene further lies in the non-volatile and reconfigurable nature of vdW polaritons[2,4]. On the practical side, the observed WGMs can be made to span a wide range of frequencies (mid-IR to THz) by adjusting the graphene carrier density in cavities with appropriate dimensions. When integrated in more complex vdW heterostructures, graphene WGMs can strongly couple with other polaritons in proximal layers, presenting a straightforward strategy for engineering versatile on-demand polaritonic responses.

Finally, we note the asymmetric spectral response of OCT cavities deep in the sub-wavelength limit ($r$ ~ 200 nm << $\lambda_p$) (**Fig. S**13). These spectral traces are reminiscent of Fano resonances in photonic nanostructures arising from the interference between a discrete oscillator and a continuum[26]. We calculated the nano-IR spectra for a discrete cavity mode interacting with a plasmonic continuum in graphene immediately surrounding the cavity (**Supplementary Discussion** 4). Our calculations account for the observations, suggesting the possibility of plasmon coupling to a broad continuum as a possible origin of the asymmetric profile.

Taken together, the data in **Figs. 1**-**4** demonstrate that our OCT-based nano-imprinting produces plasmonic cavities of exceptional quality with performance metrics limited only by the intrinsic properties of graphene. Importantly, the plasmonic signatures of whispering-gallery modes open pathways for tailored strong light-matter coupling in vdW materials integrated quantum cavities[8], plasmonic sensors[27], and cavity optomechanical systems[28]. Furthermore, the ability to produce laterally abrupt carrier density profiles using OCT provides an exciting general platform for imprinting nanoscale elements for polariton wavefront engineering and sub-wavelength lensing[29]. On the material side, we can extend our OCT approach to other families of TMOs spanning a wide range of work-functions. In particular, low work-function TMOs could serve as efficient *n*-type dopants, enabling photonic devices incorporating sharp *p-n* junctions[30]. We further remark that OCT is dictated by simple electrostatic boundary conditions, presenting a straightforward design principle to tailor charge transfer in various vdW heterostructures.

## Methods

**Heterostructure fabrication.** Our high-quality monolayer $WO_x$/graphene/hBN heterostructures are fabricated by first assembling monolayer $WSe_2$/graphene/hBN using a polymer-based dry transfer technique (**Fig. S**14). Monolayer $WO_x$ is formed by oxidizing the as-prepared heterostructures in a UV-ozone generator (Jelight UVO-cleaner) at room temperature for 30 min with an $O_2$ flow rate of 0.5L/min. While ozone molecules are the main oxidizing agents in the oxidization processes, simultaneous irradiation of UV is critical as it facilitates the room-temperature oxidation of the entire $WSe_2$ monolayer at a relatively fast time scale via the creation of local defect sites[16]. Our oxidation process also naturally self-cleans polymer residue residing on the surface of the device. Atomic force microscopy imaging shows that the resultant $WO_x$ is homogeneous with root-mean-square roughness $R_q$ of ~0.2 nm, which is comparable to $R_q$ of ~0.18 nm for bare $WSe_2$ flakes (**Fig. S**15). Programmable OCT is performed by first preparing hBN mask layers with the desired patterns using conventional e-beam lithography processes, followed by a reactive-ion etching with a gas mixture of $O_2$/$CHF_3$ (4/40 sccm) under 60 W RF power. The hBN mask is then transferred on top of $WSe_2$/graphene/hBN heterostructures, followed by UV-ozone treatment to locally trigger OCT within the openings in hBN. Freestanding $WO_x$/graphene/$WO_x$ devices are fabricated by sequentially assembling $WSe_2$, graphene, $WSe_2$, and the hBN mask layer. The heterostructure is then transferred onto pre-etched trenches in $SiO_2$/Si substrate. The trenches are etched down by ~200 nm using the same recipe for etching hBN as described above. Finally, $WO_x$ is formed by oxidizing the as-prepared heterostructures in a UV-ozone generator, as mentioned above.

**Infrared nano-imaging measurements.** Infrared nano-imaging experiments are performed using scattering-type scanning near-field optical microscopy at room temperature under ambient conditions (Neaspec). In our experiments, the AFM tip (Arrow-EFM) acts as a plasmon launcher by scattering incident light to the high in-plane momenta characteristic of graphene plasmons. We use continuous-wave tunable mid-IR quantum cascade lasers and $CO_2$ lasers as light sources, spanning laser excitation energies from 920 cm$^{-1}$ to 1400 cm$^{-1}$. A pseudoheterodyne interferometric detection technique is used together with an AFM tip-tapping frequency of 75 kHz. The near-field signal is demodulated at the third harmonic of the tapping frequency to remove the background far-field signal.

**Kelvin probe force microscopy (KFPM) measurements.** The same type of AFM tip (Arrow-EFM) is used for infrared nano-imaging and KPFM measurements. The conductive AFM tips are made of silicon with a Pt/Ir coating with a force constant of around 2.8 N/m. To accurately extract the surface potential of the OCT-doped graphene under ambient conditions, a heterodyne-KPFM method is implemented. In our experiments, the first ($\omega_1 \approx$ 75 kHz) and second ($\omega_2 \approx$ 430 kHz) cantilever resonances are utilized for topography mapping and surface potential imaging, respectively. The surface potential of our devices is referenced to the adjacent gold (**Fig. S**3). The surface potential is further calibrated using a KPFM test sample (aluminum and gold on silicon) from Bruker.

**Electrodynamics simulations.** Numerical simulations of plasmon reflection at engineered electrostatic edges in graphene and eigenmode analysis of whispering-gallery modes were performed using a wave optics module in commercial COMSOL Multiphysics software. The graphene layer was modeled as a surface current with conductivity $\sigma$ expressed within the random-phase approximation[22]. In the limits $q \ll \omega/v_F$ and $\hbar\omega \ll E_F$ (which are satisfied in our heterostructures), the conductivity $\sigma$ can be expressed as $\sigma(\omega) = e^2/\pi\hbar^2 \times E_F\tau/(1 - i\omega\tau)$, where $\tau$ is the relaxation time. The electric field patterns were calculated within the cavity by placing a tip-like point dipole source at 120 nm above graphene. The simulated $E_z$ profiles were taken 20 nm above the surface of graphene in all simulations.

### Additional information

The data that support the findings within this paper are available from the corresponding authors upon reasonable request.

### Competing financial interests

The authors declare no competing interests.

### Acknowledgments

This work was solely supported as part of Programmable Quantum Materials, an Energy Frontier Research Center funded by the U.S. Department of Energy (DOE), Office of Science, Basic Energy Sciences (BES), under award DE-SC0019443.

**Author contributions**

B.S.Y.K., A.J.S., and M.S.C contributed equally to this work. B.S.Y.K., J.C.H, and D.N.B. conceived the project and designed the experiments. B.S.Y.K. and M.S.C. fabricated the devices with assistance from A.R., A.N., and S.H.C. B.S.Y.K. and A.J.S. performed measurements with assistance from A.S.M., L.X., and Y.D. S.L. grew the $WSe_2$ crystals. Z.S. performed graphene plasmon scattering rate and Fano resonance simulations. B.S.Y.K. performed full-wave eigenmode simulations with assistance from F.L.R. and A.S.M. B.S.Y.K. and Y.S. performed Kelvin probe force microscopy measurements. A.Z. performed cross-sectional transmission electron microscopy measurements. X.X., A.J.M, P.J.S., C.R.D., J.C.H., and D.N.B. supervised the project. B.S.Y.K analyzed the data. B.S.Y.K and D.N.B co-wrote the manuscript with input from all authors.

**Corresponding authors**

Correspondence and requests for materials should be addressed to B.S.Y.K., bsk2137@columbia.edu, or D.N.B., db3056@columbia.edu.

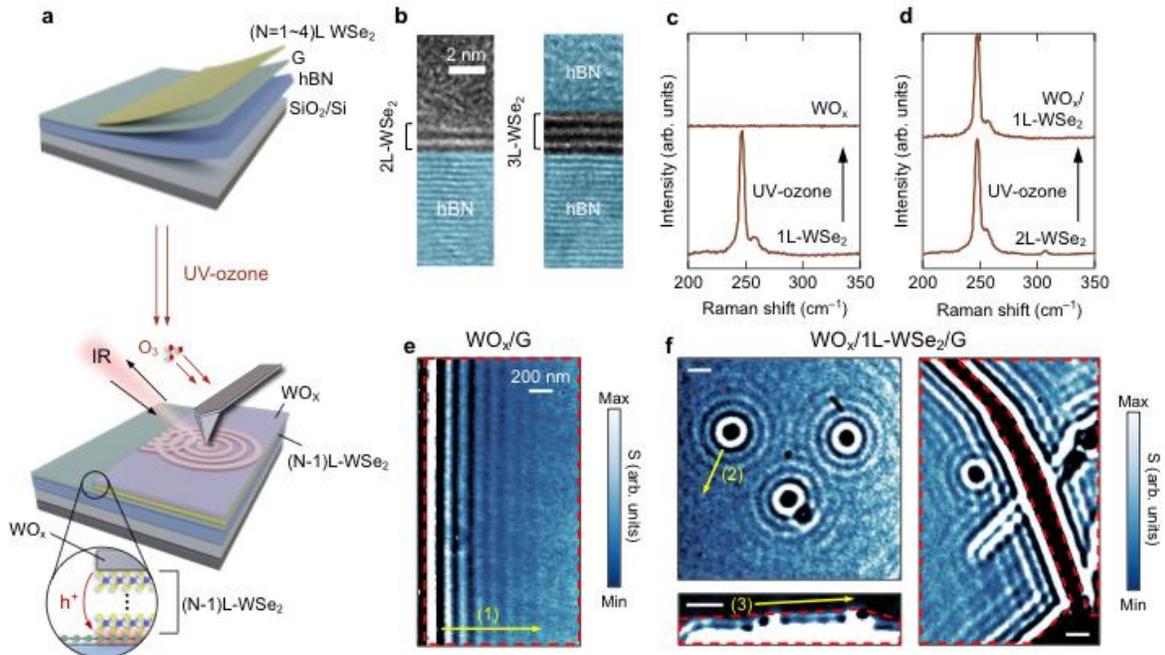

**Figure 1 | Oxidation-activated charge transfer (OCT) for on-demand graphene plasmonics. a,** Schematics of the OCT method. UV-ozone treatment of (N=1~4)L-WSe$_2$/graphene/hBN oxidizes the topmost monolayer into monolayer WO$_x$, resulting in WO$_x$/(N−1)L-WSe$_2$/graphene/hBN. The oxidation activates charge transfer between graphene and WO$_x$, which is rooted in a work-function mismatch between the two layers. **b,** Cross-sectional transmission electron microscopy (TEM) images of WO$_x$/2L-WSe$_2$/hBN (left) and hBN/3L-WSe$_2$/hBN (right). Scale bar: 2 nm. Raman spectroscopy of **c,** 1L-WSe$_2$ and **d,** 2L-WSe$_2$ before and after UV-ozone treatment. Image of nano-IR scattering amplitude $S(r,\omega)$ of **e,** WO$_x$/graphene, and **f,** WO$_x$/1L-WSe$_2$/graphene at $T = 300$ K and $\omega = 980$ cm$^{-1}$. Scale bar: 200 nm. The red dashed line denotes the boundary of WO$_x$. Within the WO$_x$/graphene regions, (1) edge- or (2) topographic defect-reflected plasmon polaritons as well as (3) one-dimensional edge modes along the WO$_x$ boundary are observed.

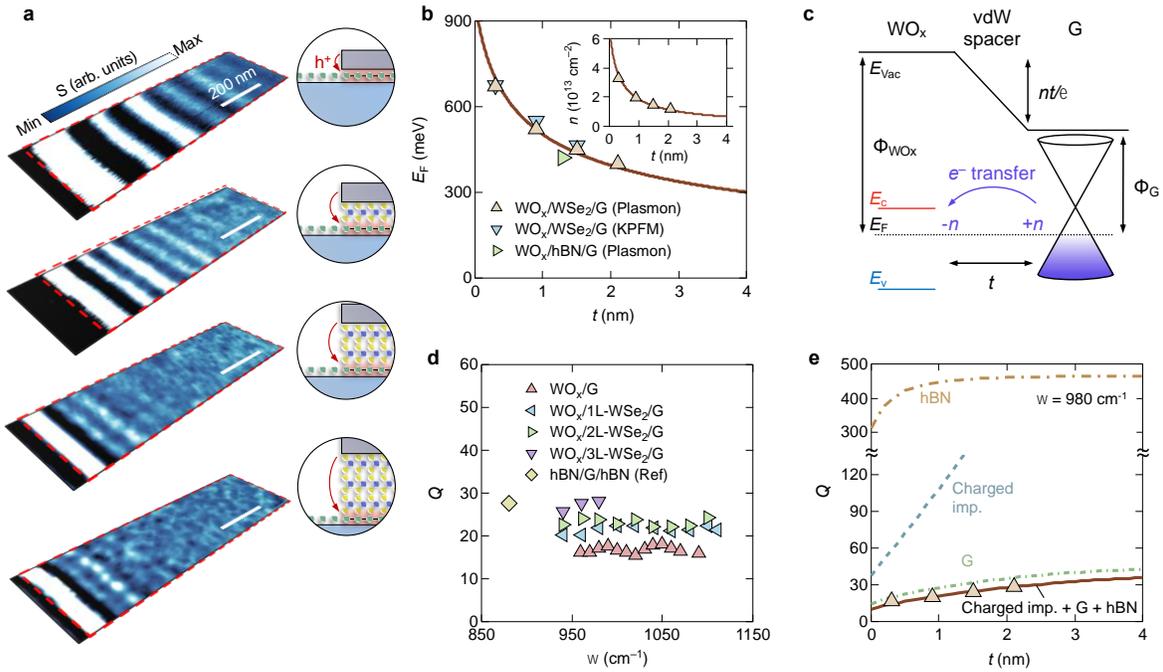

**Figure 2 | Achieving reconfigurable carrier density and high quality factor in van der Waals OCT structures. a,** Nano-IR image of the scattering amplitude signal $S(r,\omega)$ for $WO_x$/$WSe_2$/graphene/hBN heterostructures with a varied number N of $WSe_2$ spacer layers separating graphene from $WO_x$: N = 0L (top) to N = 3L (bottom). The red dashed line denotes the boundary of $WO_x$. Scale bar: 200 nm. **b,** Graphene Fermi energy $E_F$ vs spacer thickness $t$. Our experimental data obtained from near-field images (solid red triangles) and KPFM measurements (solid blue triangles) agree with the model (red solid line, see text). In our structures, $t$ is atomically tuned by varying the layer number of insulating vdW spacers ($WSe_2$ or hBN). The inset shows the corresponding carrier density $n$ as a function of $t$. **c,** Band alignment of $WO_x$-doped graphene heterostructures in equilibrium. Charge transfer occurs between graphene and $WO_x$ as a result of the work-function mismatch (see text). **d,** The quality factor $Q$ for a broad range of infrared frequencies $\omega$ and various thickness of $WSe_2$ spacers. The solid yellow diamond is $Q$ for hBN/graphene/hBN heterostructures[19]. **e,** $Q$ plotted as a function of layer thickness $t$. The solid triangles are the experimental data. The lines are the simulated $Q$ (see text) due to intrinsic phonon dissipation in graphene (green dotted line), dielectric losses of hBN (yellow dash-dotted line), charged impurities in $WO_x$ (blue dashed line), and a combination of these three key damping pathways (red solid line). An excellent agreement with the fit demonstrates that our scattering rate analysis captures the gross features of fundamental plasmonic damping pathways present in our $WO_x$-doped graphene.

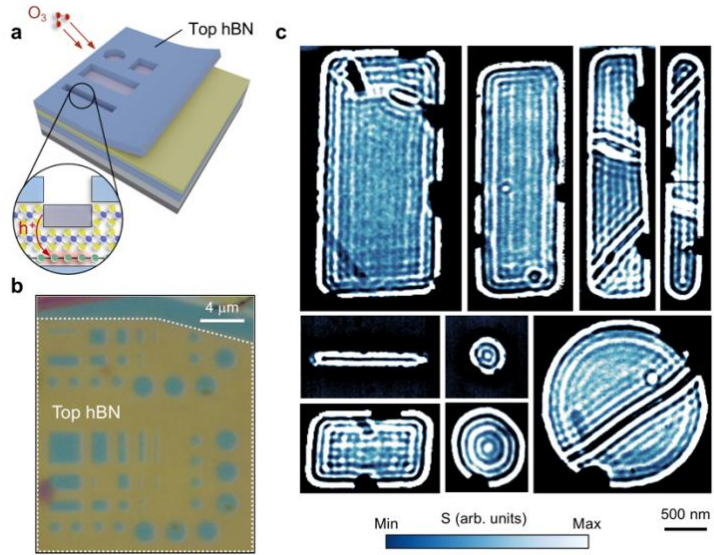

**Figure 3 | Atomically imprinted graphene plasmonic cavities.** These cavities exhibit laterally abrupt doping profiles with single-digit nanoscale precision attained via programmable OCT. **a,** Plasmonic cavities are imprinted by UV-ozone treatment in a continuous graphene monolayer using a transferred hBN mask. The OCT approach yields a laterally abrupt carrier density profile repeating the intricate patterns in the top hBN mask. **b,** Optical image of the device. Scale bar: 4 μm. **c,** Nano-IR image $S(r,\omega)$ of $WO_x$/1L-$WSe_2$/graphene heterostructures at $T = 300$ K and $\omega = 980$ cm$^{-1}$. Scale bar: 500 nm. Evident plasmonic standing-wave patterns emerge as a result of the interference between the tip-launched modes and those reflected at the engineered electrostatic edge. The presence of these fringes indicates a laterally abrupt doping profile with single-digit nanoscale precision (see text).

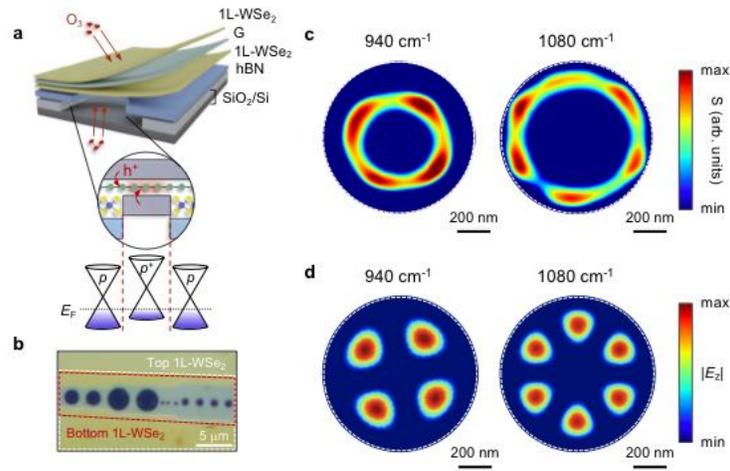

**Figure 4 | Whispering-gallery modes (WGM) in free-standing graphene plasmonic cavities encapsulated in WO$_x$. a,** Schematics of a free-standing WO$_x$/graphene/WO$_x$ plasmonic cavity embedded within moderately doped WO$_x$/graphene structures. **b,** Optical image of the device. Scale bar: 5 μm. **c,** Near-field images $S(r,\omega)$ and **d,** corresponding numerical eigenmode simulations of WGMs in cavities with a radius $r$ of ~600 nm at $\omega = 940$ cm$^{-1}$ (left) and 1080 cm$^{-1}$ (right). Scale bar: 200 nm.